\documentclass{aa}  
\usepackage{graphicx}
\usepackage{txfonts}
\begin{document} 
   \title{Structure of the equivalent Newtonian systems in MOND N-body simulations}

   \subtitle{Density profiles and the core-cusp problem}
\titlerunning{Core-cusp problem in MOND}
   \author{Federico Re
          \inst{1,2}
          \and
          Pierfrancesco Di Cintio \inst{3,4,5}
          }
   \institute{Dipartimento di Fisica ''Giuseppe Occhialini", Universit\'a di Milano Bicocca, Piazza della Scienza 3 20126, Milano, Italy
         \and
             INFN-Sezione di Milano Via Celoria 15 20133, Milano, Italy\\
          \email{federico.re@unimib.it}
          \and   
             CNR-ISC, via Madonna del Piano 17 50022 Sesto Fiorentino, Italy
             \and
             INAF-Osservatorio Astronomico di Arcetri, Largo Enrico Fermi 5 50125 Firenze Italy
             \and
             INFN-Sezione di Firenze, via Sansone 1 50022 Sesto Fiorentino, Italy\\
             \email{pierfrancesco.dicintio@cnr.it}
             }
   \date{Received ??; accepted ??}
  \abstract
   {}
   {We investigate the core-cusp problem of the $\Lambda$ cold dark matter ($\Lambda$CDM) scenario in the context of Modified Newtonian Dynamics (MOND) paradigm exploiting the concept of equivalent Newtonian system (ENS)}
   {By means of particle-mesh $N-$body simulations in MOND we explore processes of galaxy formation via cold dissipationless collapse or merging of smaller substructures. From the end states of our simulations, we recover the associated ENS and study the properties of their dark matter halos. We compare the simulation results with simple analytical estimates with a family of $\gamma-$models.}
   {We find that the dark matter density of ENSs of most spherical cold collapses have a markedly cored structure, in particular for the lowest values of the initial virial ratios. End states of some simulations with clumpy initial conditions have more complex profiles and some of their ENSs exhibit a moderate cusp, with logarithmic density slope always shallower than 1.}
   {These results seem to point towards the fact that the absence in most observed galaxies of a central DM cusp, at variance with what one would expect from theoretical and numerical arguments in $\Lambda$CDM, would be totally consistent in a MONDian description.}
   \maketitle
%
\section{Introduction}
In the $\Lambda$ cold dark matter scenario (hereafter $\Lambda$CDM), theoretical arguments and collisionless $N-$body simulations (\citealt{Navarro_etal_1997}) predict that galaxies are embedded in dark matter (DM) halos characterized by a $\rho(r)\propto r^{-1}$ central cusp. Observational results seem to suggest, from the analysis of the central velocity dispersion profiles of dwarf galaxies, that the DM distribution has a cored density distribution\footnote{Technically speaking, in multi component equilibrium self gravitating systems there exist analytical constraints on the magnitude of a component's density given the logarithmic slope of the other, e.g. see \citealt{1991ApJ...378..496D,1992MNRAS.255..561C,1996ApJ...471...68C,1999ApJ...520..574C}. Moreover for a broad range of spherical density profiles, the central density slope constraints the value of the central anisotropy profile (\citealt{2006ApJ...642..752A})} (see \citealt{1994Natur.370..629M,2014MNRAS.437..415D}).\\
\indent Several solutions to this apparent contradiction -often referred to as ``the core-cusp problem"- such as self-interacting DM (e.g. see \citealt{2012MNRAS.420.2318L,2021MPLA...3630001N,2022A&A...666A..41E}), DM annihilation (e.g. see \citealt{PhysRevD.76.103532}), baryon feedback (e.g. see \citealt{2010Natur.463..203G,10.1111/j.1365-2966.2011.19110.x,2012MNRAS.421.3464P,DelPopolo:2015nda}) or simply a misinterpretation of the observational data (\citealt{2003ApJ...584..566M}), have been proposed so far. However, notwithstanding the great amount of theoretical and observational work, a clear answer is still far from being obtained. Moreover, given the large interest in alternative theories of gravity, such as among the others $f(R)$ gravities (\citealt{1970MNRAS.150....1B,2010RvMP...82..451S}); Modified Gravity (MoG, \citealt{2006JCAP...03..004M,2013MNRAS.436.1439M}); Retarded Gravity (\citealt{2012AIPC.1483..260R,2022IJMPD..3142018Y}); Emergent Gravity (\citealt{2011JHEP...04..029V,2017ScPP....2...16V}); Refracted Gravity (\citealt{2020A&A...637A..70C,2023A&A...674A.209S}); Fractional Gravity (\citealt{2020PhRvD.101l4029G,2023ApJ...949...65B}),  proposed to avoid introducing the DM as a collisionless fluid of exotic particles, it is natural to ask what becomes of the core-cusp problem in those proposals.\\
\indent In this work we investigate this matter in the modified Newtonian dynamics (hereafter MOND, \citealt{1983ApJ...270..365M}) paradigm. We recall that in the \cite{1984ApJ...286....7B} Lagrangian formulation of MOND (sometimes referred to as AQUAL) the classical Poisson equation for a density-potential pair $(\rho;\Phi)$ 
\begin{equation}\label{poisson}
\Delta\Phi=4\pi G\rho
\end{equation}
is substituted by the non-linear field equation
\begin{equation}
\label{MOND}
   \nabla\cdot\left[\mu\left(\frac{||\nabla\Phi||}{a_0}\right)\nabla\Phi\right]=4\pi G\rho.
\end{equation}%
In the equation above where $a_0\approx 10^{-8}$cm s$^{-2}$ is a scale acceleration and $\mu(x)$ is the MOND interpolating (monotonic) function known only by its asymptotic limits
\begin{equation}
\label{limits}
\mu(x)\sim
\begin{cases}
\displaystyle 1,\quad x\gg 1,\\
\displaystyle x,\quad x\ll 1;
\end{cases}
\end{equation}%
\begin{figure}
\centering
\includegraphics[width=\columnwidth]{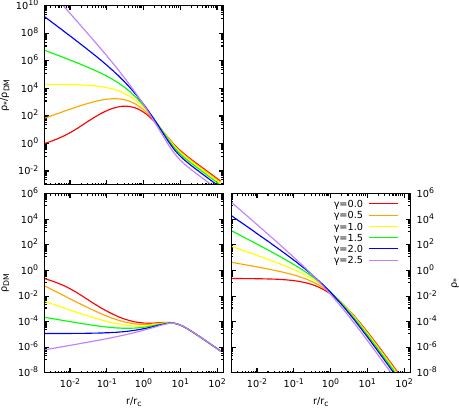}
\caption{Ratio of the stellar to dark density in the ENS (top) and DM and stellar density profiles (bottom left and bottom right) in units of $3M/4\pi r_c^3$ for $\kappa=100$ and $\gamma=0$, $0.5$, $1$, $1.5$, $2$ and $2.5$.}
\label{Logrho3}
\end{figure}
so that for $||\nabla\Phi||\gg a_0$ Eq. (\ref{MOND}) one recovers the Newtonian regime, while for $||\nabla\Phi||\ll a_0$ one obtains the so-called deep-MOND (hereafter dMOND) regime and Eq. (\ref{MOND}) simplifies to 
\begin{equation}
\label{dMOND}
   \nabla\cdot\left[||\nabla\Phi||\nabla\Phi\right]=4\pi G\rho a_0.
\end{equation}
Note that, the non-linear operator in Equation (\ref{dMOND}) is the special case of the $p-$Laplace operator (see e.g. \citealt{8bbc344b-39d3-31c4-89fe-4b604b1a5849}) for $p=3$, while Eq. (\ref{poisson}) would correspond to the $p=2$ case. In this respect, Equation (\ref{MOND}) somewhat ''interpolates" between the two regimes via the $\mu$ function. Note also that, in both cases, any given baryonic mass density $\rho$ can be taken out from Equation (\ref{poisson}) obtaining the relation
\begin{equation}
\label{gmgn}
  \mu\left(\frac{||\mathbf{g}_M||}{a_0}\right)\mathbf{g}_M=\mathbf{g}_N+\mathbf{S}
\end{equation}%
between the MOND and Newtonian force fields $\mathbf{g}_M$ and $\mathbf{g}_N$, and where $\mathbf{S}\equiv\nabla\times\mathbf{h}(\rho)$ is a density-dependent solenoidal field. It can be proved that the latter is identically null for systems in spherical, cylindrical or planar symmetry, while it is generally non-zero for arbitrary configurations of mass. On which extent the stellar system at hand with mass $M$ is dominated by MOND effects, is usually quantified by the dimensionless parameter
\begin{equation}\label{kappa}
\kappa\equiv\frac{GM}{r_c^2a_0},
\end{equation}
where $r_c$ is the scale of the baryon distribution. That is, for $\kappa\gg 1$ the system is mainly in Newtonian regime, vice versa for $\kappa\leq 1$ MOND effects become strong at all scales.\\
\indent For any given stationary model in MOND one can always build the equivalent Newtonian system (hereafter ENS), defined as a system with the same baryonic mass density $\rho_*$ plus a DM halo with density $\rho_{DM}$ such that their total potential $\Phi$ satisfying Eq. (\ref{poisson}) is the same as the MOND potential entering Eq. (\ref{MOND}) for the sole density $\rho_*$. We note that, in principle, the positivity of the DM density of the ENS is not always assured (see \citealt{1986ApJ...306....9M}), in particular for flattened systems (see \citealt{2006ApJ...640..741C,2012MNRAS.422.2058C,2016ApJ...821..111K}). We recall that \cite{2010MNRAS.403..886M} introduced a Quasi-linear formulation of MOND (hereafter QuMOND) where the modified field equation has the same form as Eq. (\ref{MOND}), with $\nu(||\mathbf{g}_N/a_0||)$ in lieu of $\nu(||\mathbf{g}_M/a_0||)$. The QuMOND interpolating function $\nu(y)$ can be recovered from $\mu(x)$ appearing in Eq. (\ref{MOND}) as 
\begin{equation}
\nu=\frac{1}{\mu}.
\end{equation}
It is easy to show that, from a given baryonic density distribution, one obtains the MONDian potential $\Phi=\Phi_N+\Phi_{pDM}$ by first solving a classical Poisson equation for the Newtonian potential $\Phi_N$, that trough an algebraic passage involving $\nu$ becomes the source for the potential $\Phi_{pDM}$ of the so-called "phantom Dark Matter" via a second application of the Poisson equation. Notably, in this alternative bi-potential MOND formulation, the DM is de facto interpreted as the effect of the second potential. As in AQUAL, in QuMOND one can retrieve a dMOND regime, that for spherical systems easily reads $g_M=\sqrt{a_0/g_N}g_N$ (see \citealt{2021PhRvD.103d4043M}).\\   
\indent If on one hand several workers have investigated the differences between static equilibrium models in Newtonian and MOND gravities, or the interpretation of observations in both theories, on the other much less is known about the formation and evolution of stellar systems. For obvious reasons, in observed systems one has access only to de-projected properties for both stellar and dark components in the Newtonian framework. Numerical experiments, though with their intrinsic limitations, yield information on the full phase-space of the simulated models; in particular the 3D density profiles. In this paper, we explore the structure of the ENS of MOND $N-$body simulations of galaxy formation, in order to shed some light on the possibility that the core-cusp problem is a MOND artifact in this paradigm of gravity. We stress the fact that the MOND core-cusp problem discussed here, is different from that introduced recently by \cite{2021A&A...656A.123E} that deals with the modified gravity versus modified inertia hypothesis (see \citealt{2022PhRvD.106f4060M}).\\
\indent The rest of this paper is structured as follows. In Sect. 2 we revise the definition of ENS and discuss their properties. In Sect. 3 we introduce the numerical models and the analysis of the simulations. In Sect. 4 we discuss the properties of the simulations' end states. Finally, in Sect. 5 we summarize and discusses the implications and the relations to previous work.
\begin{figure}
\centering
\includegraphics[width=\columnwidth]{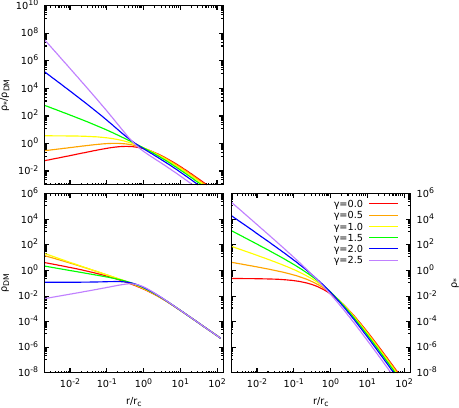}
\caption{Same as in Fig. 1 but for $\kappa=1$}
\label{Logrho4}
\end{figure}
\section{Equivalent Newtonian systems}
As anticipated above, the ENS of a MOND model is the Newtonian system with the same stellar (baryonic) mass distribution $\rho_*$ with an additional dark component $\rho_{DM}$ such that the total potential (and thus the force field) is the same of the parent MOND system (see \citealt{1994MNRAS.266..360S,2006MNRAS.371..138A}). For the case of an isolated spherical system one has
\begin{equation}\label{rhodm}
\rho_{DM}=(4\pi G)^{-1}\nabla\cdot(\mathbf{g}_M-\mathbf{g}_N),
\end{equation}
since the solenoidal term $\mathbf{S}$ vanishes. We stress the fact that, Equation (\ref{gmgn}) in QuMOND can be rewritten exactly as
\begin{equation}
\label{QuMOND}
    \mathbf{g}_M=\nu\left(\frac{||\mathbf{g}_N||}{a_0}\right)\mathbf{g}_N.
\end{equation}
If Equation (\ref{QuMOND}) is applied to a spherically symmetric system one has $\nu(y)=x/y$, and the total density of its ENS (baryonic plus phantom DM, see e.g. \citealt{2020A&A...640A..26H,2021ApJ...923...68O}) becomes
\begin{equation}
\label{ENS1}
    \rho_*(r)+\rho_{DM}(r)=\frac{d(y\nu)}{dy}\rho_*(r)-\frac{y d\nu}{dy}\frac{2}{r^3}\int_0^r \rho_*(r) r^2 dr.
\end{equation}
Let us consider the family of spherical $\gamma-$models (\citealt{1993MNRAS.265..250D,1994AJ....107..634T}), with density profile given by 
\begin{equation}\label{dehnen}
\rho_*(r)=\frac{3-\gamma}{4\pi}\frac{Mr_{c}}{r^\gamma(r+r_{c})^{4-\gamma}},
\end{equation}
where $M$ is the total baryonic mass, $0\leq\gamma<3$ is the logarithmic density slope and $r_c$ the scale radius.\\ 
\indent If the density profile (\ref{dehnen}) is substituted in Eq. (\ref{ENS1}) one obtains
\begin{equation}
\label{ENS2}
    \rho_*+\rho_{DM}=\rho_*\left[\frac{r_c}{r+r_c}\frac{d(y\nu)}{dy}-\frac{2}{3-\gamma}\frac{y d\nu}{dy}\right],
\end{equation}
where
\begin{equation}
\label{g/a0}
    y=\frac{||\mathbf{g}_N||}{a_0}=\kappa\left(\frac{r}{r_c}\right)^{1-\gamma}\left(1+\frac{r}{r_c}\right)^{\gamma-3},
\end{equation}
with $\kappa$ defined in Equation (\ref{kappa}).
We note that, for small radii $r$, Equation (\ref{g/a0}) tends to zero if $\gamma<1$, while it diverges for $\gamma>1$. In practice, at least for the $\gamma<1$ case, even in central regions the model falls in the MOND regime. Strong MOND corrections in the centre are therefore associated with a dominant DM component $\rho_{DM}$ in the ENS.
\subsection{Massive galaxies}
Let us consider a typical $10^{12}M_\odot$ massive elliptical galaxy with a scale radius of 3 kpc, modelled with a $\gamma$-model. In this case $\kappa\approx 10^2$. Due to discreteness effects of the underlying stellar system Equation (\ref{dehnen}) can be considered reliable until the radius that contains a fraction of roughly $10^{-3}$ of the total mass $M$ (in this case $10^9 M_{\odot}$, i.e. the typical mass of its central supermassive black hole). The Lagrangian radius enclosing such mass fraction is 
\begin{equation}
    r_{10^{-3}}=\frac{r_c}{10^{\frac{3}{3-\gamma}}-1}.
\end{equation}
The region in MOND regime has a far smaller radius, that for $\gamma\leq1$ is obtained by $y(r_{10^{-3}})\cong 10^5 r_{10^{-3}}^2/r_c^2$, varying between $\cong 2\times 10^2$ and $\cong 2\times 10^3$. And thus, even in the framework of (Qu)MOND the phantom DM halo does not really dominate in the central region for a cored stellar density profile. In Figure \ref{Logrho3} we plot for $\gamma=0$, 0.5, 1, 1.5, 2 and 2.5 the ratio of stellar to phantom DM and their respective radial density profiles for $\kappa=10^2$. We note that, remarkably, models with a strong cusp (i.e. $\gamma>1$) have phantom DM halos in their ENS characterized by a decreasing density inside the scale radius. Vice versa, cored models are associated with ENS having halos with a weak cusp and several slope changes. 
\subsection{Diffuse galaxies}
\indent Let us now consider a diffuse galaxy such that $\kappa\sim 1$; so that its central region can fall in the MOND regime, even for radii bigger than $r_{10^{-3}}$. Typically, this occurs again if $\gamma<1$. We find that $\lim_{r\rightarrow0}y(r)=0$ in the central region, hence $\nu(y)\sim y^{-1/2}$. If substituted in (\ref{ENS2}), this yields
\begin{equation}
    \frac{\rho_* +\rho_{DM}}{\rho_*}\sim \frac{5-\gamma}{6-2\gamma}\sqrt{\frac{a_0 r_c^2}{GM}}\left(\frac{r_c}{r}\right)^{\frac{1-\gamma}{2}};
\end{equation}
that is, the phantom DM component dominates also at small radii. In particular, the latter has a central profile given by
\begin{equation}
    \rho_{DM}\sim \frac{(5-\gamma)a_0}{8\pi Gr_c}\sqrt{\frac{a_0 r_c^2}{GM}}\left(\frac{r_c}{r}\right)^{\frac{1+\gamma}{2}}.
\end{equation}
The equation above is characterized by a weak cusp with a logarithmic density slope $\alpha=-\frac{1+\gamma}{2}>-1$. For example, for $\gamma=0$, the DM component in the ENS would have a cusp $\propto r^{-1/2}$. We note that this trend is valid for any spherically symmetric stellar distribution with a central core, and not only for the $\gamma=0$ Dehnen model. We note also it always implies a central weak cusp with logarithmic density slope $\alpha=-1/2$ for the phantom dark matter. This has the interesting astrophysical implication that a galaxy with a cored stellar density profile could be indeed interpreted in the DM scenario as having a cored halo, due to the fact that week cusps can be often mistaken for cores.\\
\indent For the cases with $\gamma>1$ and $\kappa=1$, for which the gravitational field diverges in the centre, even though the stellar density is diffuse, the ENS is DM-dominated only in the external region. This can be easily checked by substituting the asymptotic behaviour $\nu(y)\sim 1+\frac{1}{y}-\frac{1}{y^2}+o(y^{-2})$, and finding from (\ref{ENS2}) that
\begin{equation}
    \frac{\rho_{DM}}{\rho_*}\sim \frac{2}{3-\gamma}\frac{a_0 r_c^2}{GM}\left(\frac{r}{r_c}\right)^{\gamma-1} -\frac{r}{r_c} +O(r^{\gamma}),
\end{equation}
implying a vanishingly small central DM density. This is summarized in Figure 2 where we plot the same quantities as in the previous Fig. but $\kappa=1$. As expected, in the upper plot showing $\rho_*/\rho_{DM}$, the $\gamma=0$ and 0.5 (red and orange lines) are everywhere below 1, i.e. the system is dominated by the phantom DM distribution at all radii. Vice versa, for the $\gamma\geq1$ cases, the phantom DM of the ENS dominates only in the external regions. We recall that \cite{2022ApJ...940...46S} showed that galaxies with central regions in MOND regime imply ENS characterized by a decreasing baryon density and a cored DM.
\section{Numerical code and models}
\subsection{Numerical code and initial conditions}
The $N-$body simulations discussed here have been performed with a modified version of the publicly available {\sc nmody} particle-mesh MOND code (\citealt{2007ApJ...660..256N}, see also \citealt{2011ascl.soft02001L} for additional technical details). The latter uses a non-linear Poisson solver to compute $\Phi$ from Eq. (\ref{MOND}) on a $N_r\times N_\vartheta\times N_\varphi$ spherical grid in polar coordinates, using an iterative relaxation procedure starting from a guess solution (here given by Eq. \ref{gmgn} neglecting $S$), as for the linear Poisson methods (see \citealt{1990MNRAS.242..595L,1991MNRAS.250...54L}). As a rule, in the simulations discussed here we used a $128\times 32\times 64$ grid. In this work we adopt the following form for the interpolation function
\begin{equation}
\label{mu choices}
\mu(x)=\frac{x}{\sqrt{1+x^2}}.
\end{equation}
Alternative choices can also be implement always leading to qualitatively similar end states.\\
\indent The equation of motion are integrated using a standard 4$^{\rm th}$ order leapfrog scheme (see e.g. \citealt{2011EPJP..126...55D}) with an adaptive timestep $\Delta t$ conditioned by the stability threshold $\Delta t=C/\sqrt{{\rm max}|\nabla\cdot \mathbf{g}|}$, where the Courant-Friedrichs-Lewy condition $C$ was taken in the range $0.01\leq C\leq 0.1$.\\
\indent We performed two sets of numerical simulations with initial conditions defined as follows.\\
\indent In the first, the particles positions were sampled from Equation (\ref{dehnen}) while in the second, following \cite{hansen_etal}, we first distribute according to a Poissonian distribution inside a larger $\gamma$ model the centres of $N_C$ clumps\footnote{Clumpy initial conditions were also explored in the context of Newtonian simulations by \cite{2015ApJ...805L..16N} and \cite{2017MNRAS.465L..84L} when investigating the relation of the initial density fluctuation power spectrum with the S\'ersic index $m$ (see below) conjectured by \cite{2014ApJ...790L..24C}.} also described by Eq. (\ref{dehnen}) with different choices of $r_{c}$, and $\gamma$ and later populate them with particles.\\
\indent In both cases, the initial particle velocities are extracted from a position-independent isotropic Maxwell-Boltzmann distribution and normalized to obtain the wanted value of the initial virial ratio $2K/|W|$, where $K$ is the total kinetic energy and $W$ the virial function, defined for a (finite mass) continuum system of density $\rho$ and potential $\Phi$ as
\begin{equation}\label{viriale1}
W=-\int \rho(\mathbf{r})\langle\mathbf{r},\nabla\Phi\rangle d^3\mathbf{r}.
\end{equation}
We recall that in isolated dMOND systems of finite mass $W=-2\sqrt{GM^3a_0}/3$ is constant (see \citealt{2007ApJ...660..256N}). Curiously, even in systems of particles interacting with additive $1/r$ forces with logarithmic potential the virial function is constant (see \citealt{2013MNRAS.431.3177D,2017MNRAS.468.2222D}).\\
\indent The simulations of this work span a range of $N$ between $10^4$ and $10^6$. All simulations were extended up to $t=300t_{\rm Dyn}$, where $t_{\rm Dyn}\equiv\sqrt{2r_{h}^3/GM_{\rm tot}}$ and $r_h$ is the radius containing half of the total mass of the system $M_{\rm tot}$, so that virial oscillations and phase-mixing are likely to be complete.\\
\indent Following \cite{2007cpms.conf..177C} in some cases we enforce the spherical symmetry during the collapse by propagating particles only using the radial part of the evaluated force field, so that the system behaves effectively as a spherical shell model introduced in Newtonian gravity by \cite{1964AnAp...27...83H} and used in MOND among the others by \cite{2008MNRAS.386.1588S,2009ApJ...694.1220M} and by \cite{2011IJBC...21.2279D} for systems interacting with $1/r^\alpha$ forces. 
\begin{figure*}
\centering
\includegraphics[width=\textwidth]{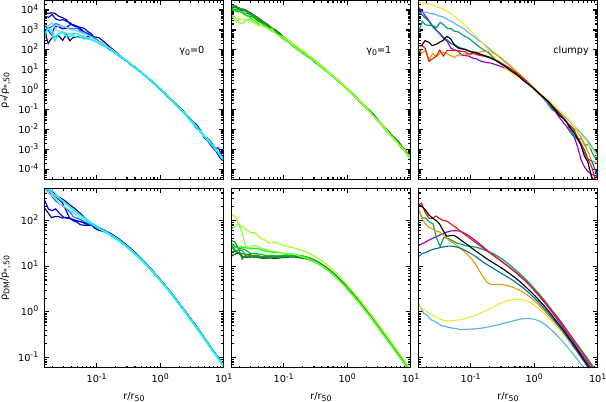}
\caption{Final ($t=300t_{\rm Dyn}$) density profiles from MOND simulations (top panels) and the DM halo  of the ENS (bottom panels) for cored $\gamma_0=0$ (left), moderately cuspy $\gamma=1$ (centre) and clumpy (right) initial conditions. The increasing initial values of the virial ratio in the models with spherical initial conditions with $\gamma=0$ and 1 is mapped with increasingly lighter tones of blue and green respectively. All clumpy initial conditions start with $2K_0/|W|_0=0.1$}
\label{rho3d}
\end{figure*}
\begin{table*}
\caption{Summary of the simulation properties: After the name of each simulation (Col. 1), we report the  number of particles (Col. 2), the gravity law (MOND or dMOND, Col. 3), the initial density profile (Col. 4), the initial virial ratio (Col. 5) the axial ratios  (Cols. 6 and 7) , the S\'ersic (Col. 8), the final anisotropy index (Col. 9)  and the virial velocity dispersion (Col. 10).} 
\begin{tabular}{lllllllllll}
\hline
Name & Gravity & Initial profile & $N$ & $2K_0/|W_0|$ & $c/a$ & $b/a$ & $m$ & $\xi$  & $\sigma_{\rm vir}$ & $\alpha$\\
\hline 
\texttt{gamma0v0}       & MOND      & $\gamma=0$   & $3\times 10^4$   & $10^{-4}$               & $0.36$ & $0.56$ & $2.45$ & $3.04$ & $0.83$ & $0.25$\\
\texttt{gamma05v0}      & MOND      & $\gamma=0.5$ & $3\times 10^4$   & $10^{-4}$               & $0.42$ & $0.62$ & $2.33$ & $2.84$ & $0.84$ & $-0.15$\\
\texttt{gamma1v0}       & MOND      & $\gamma=1$   & $3\times 10^4$   & $10^{-4}$               & $0.54$ & $0.98$ & $2.07$ & $2.73$ & $0.89$ & $-0.11$\\
\texttt{gamma15v0}      & MOND      & $\gamma=1.5$ & $3\times 10^4$   & $10^{-4}$               & $0.53$ & $0.94$ & $0.89$ & $2.91$ & $1.03$ & $-0.23$\\
\texttt{gamma2v0}       & MOND      & $\gamma=2$   & $3\times 10^4$   & $10^{-4}$               & $0.57$ & $0.92$ & $0.72$ & $2.26$ & $1.05$ & $-0.35$\\
\texttt{gamma0ve1m3}    & MOND      & $\gamma=0$   & $3\times 10^4$   & $10^{-3}$         & $0.36$ & $0.56$ & $3.34$ & $3.13$ & $0.83$ & $0.21$\\
\texttt{gamma0ve3m3}    & MOND      & $\gamma=0$   & $3\times 10^4$   & $3\times 10^{-3}$ & $0.33$ & $0.50$ & $4.28$ & $3.15$ & $0.85$ & $0.50$\\
\texttt{gamma0ve1m2}    & MOND      & $\gamma=0$   & $3\times 10^4$   & $10^{-2}$         & $0.34$ & $0.57$ & $2.53$ & $3.10$ & $0.84$ & $0.40$\\
\texttt{gamma0ve3m2}    & MOND      & $\gamma=0$   & $3\times 10^4$   & $3\times 10^{-2}$ & $0.32$ & $0.52$ & $3.06$ & $3.33$ & $0.86$ & $0.52$\\
\texttt{gamma0ve1m1}    & MOND      & $\gamma=0$   & $3\times 10^4$   & $0.1$             & $0.32$ & $0.46$ & $2.23$ & $2.98$ & $0.84$ & $0.70$\\
\texttt{gamma0ve2m1}    & MOND      & $\gamma=0$   & $3\times 10^4$   & $0.2$             & $0.33$ & $0.38$ & $3.07$ & $3.03$ & $0.83$ & $0.75$\\
\texttt{gamma0ve3m1}    & MOND      & $\gamma=0$   & $3\times 10^4$   & $0.3$             & $0.36$ & $0.37$ & $3.29$ & $3.02$ & $0.83$ & $0.85$\\
\texttt{gamma0ve4m1}    & MOND      & $\gamma=0$   & $3\times 10^4$   & $0.4$             & $0.37$ & $0.38$ & $2.91$ & $3.01$ & $0.82$ & $0.89$\\
\texttt{gamma0ve5m1}    & MOND      & $\gamma=0$   & $3\times 10^4$   & $0.5$             & $0.40$ & $0.42$ & $2.50$ & $2.92$ & $0.83$ & $1.10$\\
\texttt{gamma1ve1m3}    & MOND      & $\gamma=1$   & $3\times 10^4$   & $10^{-3}$         & $0.49$ & $0.87$ & $1.90$ & $2.86$ & $0.90$ & $-0.21$\\
\texttt{gamma1ve3m3}    & MOND      & $\gamma=1$   & $3\times 10^4$   & $3\times 10^{-3}$ & $0.50$ & $0.91$ & $3.75$ & $2.96$ & $0.91$ & $-0.14$ \\
\texttt{gamma1ve1m2}    & MOND      & $\gamma=1$   & $3\times 10^4$   & $10^{-2}$         & $0.48$ & $0.67$ & $2.48$ & $3.18$ & $0.91$ & $-0.12$\\
\texttt{gamma1ve3m2}    & MOND      & $\gamma=1$   & $3\times 10^4$   & $3\times 10^{-2}$ & $0.47$ & $0.82$ & $4.10$ & $3.20$ & $0.93$ & $-0.10$\\
\texttt{gamma1ve1m1}    & MOND      & $\gamma=1$   & $3\times 10^4$   & $0.1$             & $0.42$ & $0.62$ & $2.66$ & $3.32$ & $0.90$ & $0.00$\\
\texttt{gamma1ve2m1}    & MOND      & $\gamma=1$   & $3\times 10^4$   & $0.2$             & $0.45$ & $0.46$ & $3.10$ & $2.99$ & $0.87$ & $0.15$\\
\texttt{gamma1ve3m1}    & MOND      & $\gamma=1$   & $3\times 10^4$   & $0.3$             & $0.51$ & $0.52$ & $2.25$ & $3.33$ & $0.88$ & $0.21$\\
\texttt{gamma1ve4m1}    & MOND      & $\gamma=1$   & $3\times 10^4$   & $0.4$             & $0.54$ & $0.54$ & $2.87$ & $3.46$ & $0.88$ & $0.25$\\
\texttt{gamma1ve5m1}    & MOND      & $\gamma=1$   & $3\times 10^4$   & $0.5$             & $0.95$ & $0.96$ & $3.04$ & $3.71$ & $0.87$ & $0.45$\\
\texttt{gamma1v1em1}    & MOND      & $\gamma=1$   & $5\times 10^4$   & $0.1$             & $0.51$ & $0.95$ & $1.24$ & $2.43$ & $0.88$ & $-0.15$\\
\texttt{gamma1v0b}      & MOND      & $\gamma=1$   & $2.1\times 10^5$ & $0$               & $0.37$ & $0.69$ & $2.58$ & $2.90$ & $0.89$ & $-0.15$\\
\texttt{gamma1v0dmd}    & dMOND     & $\gamma=1$   & $2.1\times 10^5$ & $0$               & $0.24$ & $0.41$ & $2.45$ & $3.49$ & $0.82$ & $0.50$\\
\texttt{clumpy1}        & MOND      & clumpy       & $8.7\times 10^4$ & $0.1$             & $0.46$ & $0.85$ & $1.91$ & $2.72$ & $0.87$ & $0.70$\\
\texttt{clumpy2}        & MOND      & clumpy       & $8.7\times 10^4$ & $0.1$             & $0.46$ & $0.67$ & $2.86$ & $2.48$ & $0.84$ & $0.50$\\
\texttt{clumpy3}        & MOND      & clumpy       & $8.7\times 10^4$ & $0.1$             & $0.60$ & $0.97$ & $3.32$ & $1.44$ & $1.11$ & $-0.40$\\
\texttt{clumpy4}        & MOND      & clumpy       & $8.7\times 10^4$ & $0.1$             & $0.57$ & $0.67$ & $1.98$ & $2.07$ & $0.88$ & $0.05$\\
\texttt{clumpy5}        & MOND      & clumpy       & $8.7\times 10^4$ & $0.1$             & $0.66$ & $0.94$ & $3.47$ & $1.62$ & $1.18$ & $-0.75$\\
\texttt{clumpy6}        & MOND      & clumpy       & $8.7\times 10^4$ & $0.1$             & $0.41$ & $0.70$ & $1.75$ & $1.98$ & $0.85$ & $-0.80$\\
\texttt{clumpy7}        & MOND      & clumpy       & $8.7\times 10^4$ & $0.1$             & $0.30$ & $0.54$ & $1.67$ & $2.22$ & $0.83$ & $1.10$\\
\texttt{clumpy8}        & MOND      & clumpy       & $8.7\times 10^4$ & $0.1$             & $0.33$ & $0.59$ & $3.39$ & $2.02$ & $0.84$ & $0.95$\\
\texttt{clumpy9}        & MOND      & clumpy       & $8.7\times 10^4$ & $0.1$             & $0.80$ & $0.98$ & $1.64$ & $1.41$ & $1.08$ & $-0.40$\\
\texttt{clumpy10}       & MOND      & clumpy       & $8.7\times 10^4$ & $0.1$             & $0.56$ & $0.71$ & $0.98$ & $1.89$ & $0.85$ & $-0.99$\\
\texttt{clumpy1dmd}     & dMOND     & clumpy       & $8.7\times 10^4$ & $0.1$             & $0.13$ & $0.28$ & $3.46$ & $7.45$ & $0.77$ & $1.01$\\
\texttt{clumpy2dmd}     & dMOND     & clumpy       & $8.7\times 10^4$ & $0.1$             & $0.36$ & $0.61$ & $3.16$ & $2.69$ & $0.83$ & $1.00$\\
\texttt{clumpy3dmd}     & dMOND     & clumpy       & $8.7\times 10^4$ & $0.1$             & $0.35$ & $0.38$ & $1.40$ & $1.94$ & $0.82$ & $-1.99$\\
\texttt{clumpy4dmd}     & dMOND     & clumpy       & $8.7\times 10^4$ & $0.1$             & $0.30$ & $0.37$ & $1.74$ & $2.34$ & $0.81$ & $0.70$\\
\texttt{clumpy5dmd}     & dMOND     & clumpy       & $8.7\times 10^4$ & $0.1$             & $0.55$ & $0.56$ & $0.82$ & $2.08$ & $0.82$ & $-0.50$\\
\texttt{clumpy6dmd}     & dMOND     & clumpy       & $8.7\times 10^4$ & $0.1$             & $0.33$ & $0.62$ & $1.55$ & $2.00$ & $0.83$ & $0.90$\\
\texttt{clumpy7dmd}     & dMOND     & clumpy       & $8.7\times 10^4$ & $0.1$             & $0.26$ & $0.57$ & $1.44$ & $2.24$ & $0.82$ & $0.91$\\
\texttt{clumpy8dmd}     & dMOND     & clumpy       & $8.7\times 10^4$ & $0.1$             & $0.29$ & $0.59$ & $1.89$ & $2.00$ & $0.81$ & $0.99$\\
\texttt{clumpy9dmd}     & dMOND     & clumpy       & $8.7\times 10^4$ & $0.1$             & $0.35$ & $0.39$ & $1.56$ & $2.10$ & $0.81$ & $0.80$\\
\texttt{clumpy10dmd}    & dMOND     & clumpy       & $8.7\times 10^4$ & $0.1$             & $0.42$ & $0.60$ & $1.76$ & $1.70$ & $0.75$ & $0.61$\\
\texttt{gamma0v5em51D}  & MOND (1D) & $\gamma=0$   & $3\times 10^4$   & $10^{-4}$         & $0.96$ & $0.98$ & $2.26$ & $20.5$ & $0.71$ & $-0.01$\\
\texttt{gamma05v5em51D} & MOND (1D) & $\gamma=0.5$ & $3\times 10^4$   & $10^{-4}$         & $0.98$ & $0.99$ & $2.27$ & $33.1$ & $0.98$ & $-0.45$\\
\texttt{gamma1v5em51D}  & MOND (1D) & $\gamma=1$   & $3\times 10^4$   & $10^{-4}$         & $0.97$ & $0.97$ & $1.62$ & $88.0$ & $1.63$ & $-0.40$\\
\texttt{gamma15v5em51D} & MOND (1D) & $\gamma=1.5$ & $3\times 10^4$   & $10^{-4}$         & $0.97$ & $0.99$ & $0.71$ & $44.8$ & $1.48$ & $-0.21$\\
\texttt{gamma2v5em51D}  & MOND (1D) & $\gamma=2$   & $3\times 10^4$   & $10^{-4}$         & $0.96$ & $0.97$ & $0.51$ & $451$ & $4.72$  & $-0.35$\\
\texttt{gamma1v3em31D}  & MOND (1D) & $\gamma=1$   & $3\times 10^4$   & $3\times 10^{-3}$ & $0.97$ & $1.00$ & $2.10$ & $4.47$ & $0.66$ & $0.01$\\
\texttt{gamma1v1em21D}  & MOND (1D) & $\gamma=1$   & $3\times 10^4$   & $10^{-2}$         & $0.97$ & $0.99$ & $3.78$ & $15.8$ & $0.60$ & $-0.10$\\
\texttt{gamma1v3em21D}  & MOND (1D) & $\gamma=1$   & $3\times 10^4$   & $3\times 10^{-2}$ & $0.96$ & $0.98$ & $3.11$ & $5.19$ & $0.63$ & $-0.51$\\
\texttt{gamma1v1em11D}  & MOND (1D) & $\gamma=1$   & $3\times 10^4$   & $0.1$             & $0.97$ & $0.99$ & $2.42$ & $4.02$ & $0.64$ & $-0.62$\\
\texttt{gamma1v2em11D}  & MOND (1D) & $\gamma=1$   & $3\times 10^4$   & $0.2$             & $0.98$ & $0.99$ & $3.61$ & $3.19$ & $0.62$ & $-0.45$\\
\texttt{gamma1v3em11D}  & MOND (1D) & $\gamma=1$   & $3\times 10^4$   & $0.3$             & $0.99$ & $0.99$ & $4.09$ & $2.57$ & $0.63$ & $-0.35$\\
\hline
\end{tabular}
\label{tab_res}
\end{table*}
\subsection{Analysis of the end products}
For all simulations presented here we first extract the intrinsic properties of the end products from their phase-space positions. We first evaluate the triaxiality of the final particle distribution (see e.g. \citealt{2006MNRAS.370..681N,2013MNRAS.431.3177D} and references therein) by defining the tensor 
\begin{equation}
I_{ij}\equiv m\sum_{k=1}^N r_i^{(k)}r_j^{(k)} 
\end{equation}
for the particles with positions $\mathbf{r}_i$ within the Lagrangian radius $r_{70}$ containing the 70\% of the stellar mass of the system and evaluating with a standard iterative procedure its three eigenvalues $I_1\geq I_2\geq I_3$.  By applying a rotation $\mathcal{R}$ to all particles of the system so that the three associated eigenvectors are now oriented along the coordinate axes we then get that the three semiaxes $a\geq b\geq c$ from $I_1=Aa^2$, $I_2=Ab^2$ and $I_3=Ac^2$, 
where $A$ is a numerical constant depending on the density profile. Finally, we define the axial ratios $b/a=\sqrt{I_2/I_1}$ and $c/a=\sqrt{I_3/I_1}$, and the ellipticities in the principal planes $\epsilon_1=1-\sqrt{I_2/I_1}$ and $\epsilon_2=1-\sqrt{I_3/I_1}$.\\
\indent Following \cite{2007ApJ...660..256N} and \cite{2013MNRAS.431.3177D} we compare the surface density profiles of the end products with the \cite{1968adga.book.....S} law
\begin{equation}\label{sersic}
\Sigma(R)=\Sigma_e e^{-b\left[\left(\frac{R}{R_e}\right)^{1/m}-1\right]},
\end{equation}
where $\Sigma_e$ is the projected mass density at effective radius $R_e$, the radius of the circle containing half of the projected mass, and the dimensionless parameters $b,m$ are related by $b\simeq2m-1/3+4/405m$ as found by \cite{1999A&A...352..447C}.\\
\indent Once the projected density in the 3 principal planes is circularized over elliptical shells, we determine the corresponding pair $(R_e,\Sigma_e)$ by particle counts (i.e. we are assuming a constant mass to light ratio for each particle), and fit Eq. (\ref{sersic}) for the three projections. We find that, in general, all 3 sets of ($\Sigma_e$, $R_e$, $m$) are rather similar (differing only for less than the 5\%), we therefore chose randomly only one.\\
\indent In addition, for all simulations we also evaluate the so-called anisotropy index (see \citealt{2008gady.book.....B}) defined by
\begin{equation}
\xi=\frac{2K_r}{K_t},
\end{equation}
where $K_r$ and $K_t=K_\theta+K_\phi$ are the radial and tangential components of the kinetic energy tensor, respectively and read
\begin{equation}
K_r=2\pi\int\rho(r)\sigma^2_r(r)r^2{\rm d}r,\quad K_t=2\pi\int\rho(r)\sigma^2_t(r)r^2{\rm d}r.
\end{equation} 
In the expressions above, $\sigma^2_r$ and $\sigma^2_t$ are the radial and tangential phase-space averaged square velocity components and are obtained for the end products of the simulations by particle counts over radial shells.\\
\indent For each simulation we recover the (spherical) DM density of the ENS from Equation (\ref{rhodm}) where the Newtonian force field $g_N$ has been evaluated and averaged on the radial coordinate. In practice, we are assuming a ''sphericized" system. 
\begin{figure}
\centering
\includegraphics[width=0.85\columnwidth]{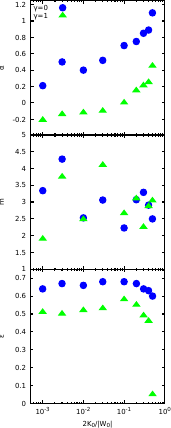}
\caption{Inner density slope of the ENS halo (top panel), best fit S\'ersic
index (middle panel) and minor ellipticity $\epsilon=1-c/a$ as function of the
initial virial ratio for initial conditions with Dehnen profiles with $\gamma=0$ (circles) and 1 (triangles)}
\label{alfa}
\end{figure}
Finally, for the density distribution $\rho_{DM}$ so obtained we evaluate the logarithmic density slope $\alpha$. We find that the profiles of $\rho_{DM}$ are generally well fitted by the empirical law
\begin{equation}\label{rhoempiric}
\rho(r)=\frac{\rho_\alpha r_\alpha^2}{r^{\alpha}(r^2+r_\alpha^2)^{\frac{2-\alpha}{2}}},
\end{equation}
where $r_{\alpha}$ is a scale radius and $\rho_{\alpha}$ is the associated scale density. Equation (\ref{rhoempiric}) above recovers the $1/r^2$ trend of the density of the ENS as predicted by the logarithmic behaviour of the far field MOND potential. The properties of the simulations and their initial conditions are summarized in Tab. \ref{tab_res} below.
\section{$N-$body simulations}
\subsection{Spherical collapses}
One of the main motivations of the present work is to establish whether the end products of MOND dissipationless collapses could, in principle, reproduce the structural properties of elliptical galaxies together with their inferred dark halos. Single component Newtonian collapses with spherical initial conditions, are known to produce flatter end states for increasing values of their initial virial ratio (see \citealt{2006MNRAS.370..681N,2006ssc..rept..122N,2013MNRAS.431.3177D} and references therein) at fixed initial density profile.\\
\indent We find that, this (partially) holds true for MOND spherical collapses, as shown in Fig. \ref{rho3d} (top left and top mid panels), where we plot the baryon density distribution at $300t_{\rm Dyn}$ for $\gamma=0$ and $1$ and increasing values of the virial ratio with increasingly lighter tones of blue and green in the range $10^{-3}\leq2K_0/|W_0|\leq 0.5$. Using Equation (\ref{rhodm}) for the angle-averaged final density profile on a spherical grid, we evaluated the density distribution of the DM component of the parent Newtonian model (see bottom panels, same figure). We find that, in qualitative agreement with the structural properties of the ENS (see Figs. \ref{Logrho3} and \ref{Logrho4} in Sect. 2), cuspy end systems can be associated with cored or weakly cuspy phantom halos. In general, the end products of spherical collapses have always inner regions that are baryon dominated when building their ENS, even if the initial conditions are such that $\kappa=1$ (in particular for the $\gamma=0$ cases).\\     
\indent Consistently with \cite{2007ApJ...660..256N}, we observe that, independently on the specific value of the initial virial ratio, initial conditions characterized by a moderate density cusp (i.e. $0.5\leq\gamma\leq 2$) tend to yield end products that are in general oblate (i.e. $0.5\lesssim c/a\lesssim b/a$), as for Newtonian single component collapses. We typically observe major ellipticities up to $\sim 0.63$ (corresponding to the \texttt{gamma1v0b} case, see Tab. \ref{tab_res}). Remarkably, MOND collapses with cored initial conditions (i.e. $\gamma=0$) evolve into rather prolate end states for $2K_0/|W_0|\gtrsim 0.1$, and markedly triaxial end states for lower values of the initial virial ratio. For both cored and moderately cuspy initial conditions, the inner slope $\alpha$ of the DM halo of the ENS, obtained by fitting with Eq. (\ref{rhoempiric}) increases for increasing values of the baryon initial virial ratio in the MOND simulation, as shown in 
 Fig. \ref{alfa} (top panel). The best fit S\'ersic index $m$, measuring the concentration of the projected stellar density profile is always in the range $2\leq m\leq 4.5$ for both choices of the initial density profile (mid panel, same figure), while the major ellipticity $\epsilon=1-c/a$ is typically larger when the initial condition has a lower virial ratio, being smaller for larger values of the initial $\gamma$ at fixed $2K_0/|W_0|$ (bottom panel, same figure). Remarkably no system is found being more flattened than an E7 galaxy. However, as also found by \cite{2007ApJ...660..256N}, dMOND collapses may produce even flatter end states as in the case of the \texttt{gamma1v0dmd} run, for which $c/a\sim 0.24$ so that $\epsilon=0.76$. For fixed initial virial ratio, the end states attain larger values of the central virial velocity dispersion $\sigma_{\rm vir}$ for increasing values of the initial density slope, while the anisotropy index $\xi$ decreases (cfr. \ref{tab_res}). At fixed initial density profile, the final values of $\sigma_{\rm vir}$ have little variation with $2K_0/|W_0|$, while $\xi$ is usually lower for the relaxed states of hotter initial conditions.\\
 \indent In order to clarify whether the properties of the halo in the ENS are or not an artifact of the angle averaging procedure, we also performed a set of simulation in enforced spherical symmetry by propagating particles only using the radial component of the force field. By doing so, the system remains spherically symmetric (as no radial orbit instability is possible) and $S=0$ de 
\begin{figure}
\centering
\includegraphics[width=0.85\columnwidth]{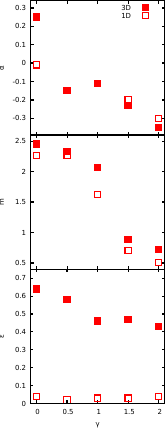}
\caption{Inner density slope of the ENS halo (top panel), best fit S\'ersic index (middle panel) and minor ellipticity $\epsilon=1-c/a$ as function of the logarithmic density slope $\gamma$ of the initial condition for full 3D (filled symbols) and 1D simulations (empty symbols).}
\label{alfag}
\end{figure}
 facto hold true at all times, so that one could apply Eq. (\ref{rhodm}) exactly. In Figure \ref{alfag} we show the same quantities as in Fig. \ref{alfa} as function of the initial values of $\gamma$ for systems starting with a virial ratio of $10^{-4}$ with (empty symbols) and without (filled symbols) enforced spherical symmetry. The trend as well as the values of $\alpha$ 3D and effective 1D simulations are comparable, and the same could be noted also for the S\'ersic index $m$ that attains considerably lower values (associated with a more concentrated density profile) for larger values of the initial logarithmic density slope. In all cases (cfr. \ref{tab_res}), as expected, 1D collapses relax to final stated with rather large values of the orbital anisotropy $\xi$.\\
\indent Figure \ref{3dvs1d} shows the final angle averaged density profiles for $\gamma_0=0$, 1 and 1.5 in 3D and 1D simulations (solid lines) as well as the density profiles of the ENS halos (dashed lines). Notably, if on one hand the large $r$ behaviour of the baryon density profiles $\rho_*$ (where the systems are mostly dominated by radial orbits) does not change significantly, on the other, the inner slope of $\rho_*$ is always higher for the end products of the 1D simulations and typically settles around $2.5$. With the sole exception of the cored initial conditions ($\gamma_0=0$), the DM halo of the ENS of the end products is denser (in units of the baryon component density $\rho_{*,50}$ evaluated at the half mass radius $r_{50}$) for the 1D simulations, being in both cases considerably shallower than the parent baryon density.
\begin{figure*}
\centering
\includegraphics[width=0.85\textwidth]{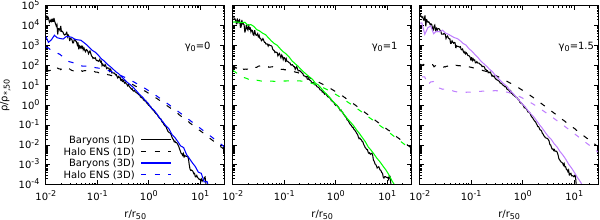}
\caption{Final baryon density profiles (coloured solid lines) and ENS halos (coloured dashed lines) for $\gamma=0,$ 1 and 1.5. The black lines refer to the 1D the cases with the same initial conditions.}
\label{3dvs1d}
\end{figure*}
\begin{figure}
\centering
\includegraphics[width=\columnwidth]{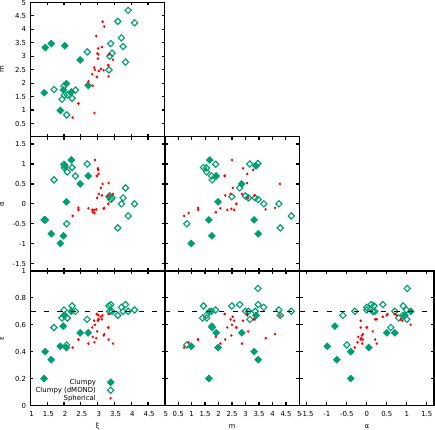}
\caption{Matrix plot of the S\'ersic index, slope of the DM profile in the ENS, major ellipticity and anisotropy index for  simulations with clumpy (diamonds) and spherical (circles) initial conditions. Empty symbol mark the dMOND runs.}
\label{clumpy}
\end{figure}
\begin{figure}
\centering
\includegraphics[width=0.9\columnwidth]{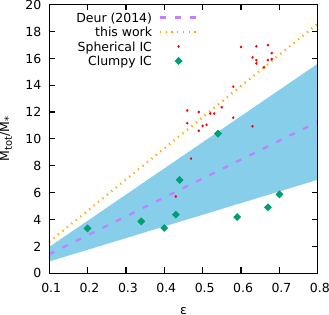}
\caption{Mass ratio against ellipticity relation for the final states of spherical (red circles) and clumpy (green diamonds) initial conditions. The purple dashed line marks the \cite{2014MNRAS.438.1535D} relation with its uncertainty (blue shaded area), while the orange dotted line marks the linear fit for the models with spherical initial conditions.}
\label{epsy}
\end{figure}
\subsection{Clumpy collapses}
The numerical studies in MOND carried out so far, have typically explored spherical initial conditions (see \citealt{2007ApJ...660..256N,2007cpms.conf..177C,2008MNRAS.386.1588S,2009ApJ...694.1220M,2011MNRAS.414.3298N}), disks (\citealt{1999ApJ...519..590B,2007A&A...464..517T,2008A&A...483..719T,2007MNRAS.379..597N,2017MNRAS.468.4450G,2020ApJ...890..173W})  or galaxy merging (\citealt{2007MNRAS.381L.104N,2008ASPC..396..259T}) and references therein. Here, in addition to the usual spherical collapses we also explored clumpy initial conditions. When starting with such initial states, MOND simulations tend (as expected) to yield markedly triaxial end states with broader ranges of both $c/a$ and $b/a$. In general, for fixed values of the initial virial ratio, the systems tend to relax at later times with respect to their initially spherical counterparts (the oscillations of $2K/W$ damp out at about $50t_{\rm Dyn}$ in spherical collapses, see \citealt{2007ApJ...660..256N}, while in clumpy systems this happens on average at around $140t_{\rm Dyn}$) for analogous choices of the virial ratio. We report here only the runs corresponding to $2K_0/|W_0|=0.1$, see Tab. \ref{tab_res}.\\
\indent The final three dimensional (angle averaged) density profiles (see top right panel in Fig. \ref{rho3d}) are strikingly more complex than those obtained from spherical initial conditions and bare individually more slope changes. The projected 2D density profiles are fitted by the S\'ersic law with roughly the same (percentage) asymptotic standard error of about, on average, 3\% as for the spherical collapses, while the scatter in the $m$ S\'ersic parameter is slightly smaller for MOND clumpy systems (see top panel in Fig. \ref{clumpy}). For comparison, we also run the same clumpy initial conditions in dMOND finding a larger scatter in $m$.\\
\indent As a general trend, the DM halo of the circularized ENS of clumpy collapses are significantly more cored\footnote{Notably, 
 in Newtonian simulations of clumps in fall in DM halo \cite{10.1111/j.1365-2966.2011.19110.x} found that the central DM cusp is considerably weakened by the collapsing clumpy satellites.} than what is typically obtained in spherical collapses. In several cases  the inner density slopes are negative, down to $\sim-0.99$, corresponding to a DM density profile that decreases in the central regions (middle panels in Fig. \ref{clumpy}). Interestingly, no initially clumpy system is found to evolve into a state flatter than an E7 galaxy (thin dashed line in bottom panels of Fig. \ref{clumpy}) in MOND simulations. However, some dMOND collapses result in considerably flatter end states (and often prolate) with major ellipticity reaching 0.87 for the \texttt{clumpy1dmd}.\\
\indent We observe that, final states with larger values of the anisotropy index $\xi$ (i.e. more and more dominated by low-angular momentum orbits) are always associated to larger ellipticities $\epsilon$ and S\'ersic indexes. A similar, though somewhat weaker, correlation is also found between $\alpha$ and $\epsilon$, that could be read in the DM scenario as steeper inner DM profiles producing flatter stellar distributions. 
\subsection{The MOND mass-to-light ratio - ellipticity relation}
\cite{2014MNRAS.438.1535D,2020arXiv201006692D} and more recently \cite{2023MNRAS.518.2845W}, using a broad sample of elliptical galaxies from independent surveys, and different methods to evaluate the mass to light ratio $M/L$ (i.e. Jeans anisotropic modelling, gravitational lensing, X-ray spectra and the dynamics of satellite star clusters) and the ellipticity $\epsilon$, 
recovered the linear relation
\begin{equation}\label{epstoml}
  M/L = (14.1 \pm 5.4)\epsilon, 
\end{equation}
where the $M/L$ is normalized such that $M/L(\epsilon_{\rm app}=0.3)\equiv 8M_{\odot}/L_{\odot}\equiv 4M/M_*(\epsilon_{\rm app}=0.3)$, and the intrinsic ellipticity $\epsilon$ is extrapolated from its observed 2D projected value $\epsilon_{\rm app}$ assuming that all systems are oblate with a Gaussian distribution of projection angles $\theta$ so that
\begin{equation}\label{epsilonapp}
\epsilon_{\rm app}=1-\sqrt{(1-\epsilon)^2\sin^2\theta+\cos^2\theta}.
\end{equation}
The Equation above in the context of $\Lambda$CDM implies that a larger contribution of the DM mass $M_{\rm DM}$ to the total mass $M$ corresponds to a larger departure from the spherical symmetry (here quantified by larger major ellipticity) for the stellar component. \cite{2023MNRAS.518.2845W} argue that, if true, such a correlation would be contrasting the standard $\Lambda$CDM scenario of galaxy formation, where more massive (and rather spherical) DM halos are embed less flattened stellar systems. We note that some peculiar elliptical galaxies (though excluded by the original sample of \citealt{2023MNRAS.518.2845W}) such as the ultrafaint dwarfs (\citealt{2019ARA&A..57..375S}) appear to go against the trend given by Eq. (\ref{epstoml}), having usually $\epsilon\lesssim 0.1$ with $M/L$ in some cases up to $10^3$.\\
\indent Using the simulations discussed in the previous sections, we have investigated the relation (\ref{epstoml}) in the context of MOND, evaluating the effective DM mass $M_{\rm DM}$ in the ENSs of both clumpy and spherical collapses. To do so, after recovering the $\rho_{DM}$ from the angle averaged ENS, we integrate it radially up to the radius containing all simulation particles.\\
\indent In Figure \ref{epsy} we show the total to stellar mass (here we have assumed units such that $M_*/L=1$) ratio $M/M_*$ versus major ellipticity $\epsilon$ for collapses with both spherical and clumpy initial states, here indicated by circles and diamonds respectively, as well as the observational relation given in Eq. (\ref{epstoml}). We found that the end products of initially clumpy systems fall in (almost) all cases within \cite{2023MNRAS.518.2845W}'s relation and its error range (indicated in figure by the shaded area), while for the spherical collapses fall on a rather steeper relation. We performed a linear fit (marked in figure by the orange dotted line) obtaining
\begin{equation}\label{epstoml2}
  M/M_* = (23.24 \pm 0.59)\epsilon.
\end{equation}
We stress the fact, that none of the simulations discussed above produces final states that could be interpreted as ultrafaint dwarfs (except, possibly, some dMOND collapses), that in the standard cosmological scenario are supposed to be DM dominated at all radii (i.e. even in the central region where our simulations, when interpreted in the context of DM have baryon dominated cores).
\section{Discussion and conclusions}
In this work we have investigated the structure of the dark matter density profiles of the (angular averaged) equivalent Newtonian systems of the end states of MOND dissipationless collapse simulations. We studied a broader range of initial conditions than those discussed by \cite{2007ApJ...660..256N}, including non spherical ones.\\
\indent The main results of this work can be summarized as follows:
Simple analytical estimates in spherical symmetry suggest that the presence of a core or even centrally decreasing DM distribution in ENS of MOND models with cuspy stellar profiles. Vice versa, cored stellar profiles are associated 
 with ENS DM central density profiles $\rho_{DM}\propto1/r^{\alpha}$ with $\alpha\lesssim 1$. Our MOND $N-$body simulations and the angle averaged ENS of their end states nicely confirm this. This established, we can conclude that he flat-cored halos invoked by some observational studies, can be reasonably considered in agreement with our numerical finding, as the dynamical effect of a weak cusp, independently on the specific value of the central logarithmic density slope of the baryons, can be easily mistaken for that of a cored dark mass distribution in the DM paradigm.\\
 \indent In general, we observe that as for the simulations in Newtonian gravity, in MOND the stronger is the collapse (i.e. lower initial virial ratio and/or larger initial density slope), the steeper is the final density profile, and thus the dark halo of the ENS has a markedly cored, or sometimes even depleted, inner density. Obviously, the end product of simplified MOND $N-$body simulations with enforced spherical symmetry have ENS with markedly flat cores, for a broad spectrum of initial values of density slope and virial ratio, with baryon density always dominated by a rather strong cusp at inner radii. Moreover, we also find that, if interpreted in the context of DM, the relaxed end states with smaller values of the ellipticity (i.e. less flattened) should have cuspier DM halos. In general, independently on the specific form of the initial density profile, colder initial conditions are always associated to flatter end states.\\
 \indent As a by-product of this simulation study on ENSs, we have also recovered a numerical confirmation of the claimed \cite{2014MNRAS.438.1535D} observational linear correlation between $M/L$ (or $M/M_*$) and $\epsilon$, though with seemingly different slope, when evaluating the dark matter content of ENSs in units of the baryon mass (the latter being a pre-defined simulation parameter).\\
\indent Our findings lead us to speculate that in the context of MOND the core cusp problem could be a ``MOND artifact" in the same sense as rings and DM shells discussed by \cite{2008ApJ...678..131M}. Moreover, we stress the fact that in the DM halos reconstructed from observational data using the line-of-sight velocity dispersion of a given tracer stellar population, the effect of the velocity anisotropy profiles $\beta(r)=1-\sigma_t^2(r)/2\sigma_r^2(r)$ (and the intrinsic departure from the spherical symmetry) is neglected, as noted by \cite{2009MNRAS.393L..50E} for the case of dwarf spheroids. In fact, since the central stellar $\beta$ profile imposes a constraint on the slope of the DM component in the form of the inequality $\beta\leq\alpha/2$ (see \citealt{2006ApJ...642..752A,2009MNRAS.393..179C,2010MNRAS.408.1070C}) the entity of the central density cusp or core inferred for observed galaxies is likely to bare a rather large uncertainty. In the context of (single component) MOND models, the relation between anisotropy and central density cusps has not been explored in detail, neither analytically nor in simulations. Simple numerical experiments (\citealt{2013MNRAS.431.3177D}) with inverse power-law radial forces seem to suggest that the density slope anisotropy inequality is a rather general property of the relaxed states of collapses with long-range interactions.\\
\indent A natural follow up of this work will be a systematic study of the interplay of the $\beta$ profiles in MOND systems and the DM density profiles of the parent ENSs. 
\begin{acknowledgements}
  We would like to express gratitude to Carlo Nipoti for the assistance with the simulation in {\sc nmody} and Michal B\'ilek for the discussions at an early stage of this work. One of us (PFDC) wishes to acknowledge funding by ‘‘Fondazione Cassa di Risparmio di Firenze" under the project {\it HIPERCRHEL} for the use of high performance computing resources at the university of Firenze.
\end{acknowledgements}
   \bibliographystyle{aa} 
   \bibliography{biblio} 
\end{document}